\documentclass[aps,prb,twocolumn,10pt,longbibliography,floatfix,superscriptaddress]{revtex4-2}

\usepackage{physics}
\usepackage[english]{babel}
\usepackage[utf8]{inputenc}
\usepackage{amssymb}
\usepackage{amsmath}
\usepackage{dsfont}
\usepackage{multirow}

\usepackage[pdftex]{graphicx}
\usepackage{xspace}
\usepackage{dsfont}
\usepackage{bm}
\usepackage[pdfencoding=auto, psdextra]{hyperref}
\hypersetup{
    colorlinks,%
    citecolor=blue,%
    filecolor=blue,%
    linkcolor=blue,%
    urlcolor=blue
}

\usepackage[usenames, dvipsnames]{xcolor}

\graphicspath{ {Fig/} }
\providecommand{\mean}[1]{\expval{#1}}

\begin{document}

\title{Fabry-Pérot resonant vortices and magnetoconductance \\ in topological insulator constrictions with magnetic barriers}
\author{R.~P.~Maciel}
\author{A.~L.~Araújo}
\affiliation{Instituto de Física, Universidade Federal de Uberlândia, Uberlândia, 38400-902 MG, Brazil}
\author{C.~H.~Lewenkopf}
\affiliation{Instituto de Física, Universidade Federal Fluminense, 24210-346 Niterói, RJ, Brazil}
\author{G.~J.~Ferreira}
\affiliation{Instituto de Física, Universidade Federal de Uberlândia, Uberlândia, MG 38400-902, Brazil}
\date{\today}

\begin{abstract}
The edge states of two-dimensional time-reversal topological insulators support a perfect helical conductance on wide ribbons due to the absence of backscattering. Here, we study the changes in the transport properties of topological insulator nanoribbons by introducing a constriction along the ribbon. This set up allows the edge states to hybridize, leading to reflections at the ends of the constriction. We find that the electronic states running along one edge can be reflected back along the opposite edge multiple times, giving rise to Fabry-Pérot resonant vortices within the constriction with well defined conductance peaks. We show that magnetic barriers allow one to manipulate these peaks and obtain significant changes in the system spin-resolved magnetoconductance.
\end{abstract}

\maketitle




\section{Introduction}

The edge states of two-dimensional (2D) topological insulators (TIs) promote a helical dynamics around the crystal, which is dictated by massless Dirac-like Hamiltonians \cite{bernevig2006quantum, Bernevig2006BHZ, moore2010birth, hasan2010TIColloquium, liu2010model, qi2011topological, Kou2017TI2DProgress, Vergniory2019}. Consequently, time-reversal symmetry enforces the absence of backscattering, since a change in direction requires a spin flip. The quantized conductance of these 1D channels has been observed in HgTe quantum wells \cite{Konig2007QSHIHgTe, Gusev2014}, including an anomalous 0.5-plateau possibly due to strong correlations \cite{Strunz2020Anomaly05}, while the corresponding Dirac spectrum has been directly measured by ARPES \cite{Xia2009, Zhang2010, Wang2011, Landolt2014, Dil2019ARPES}. The spin-momentum locking of these states leads to promising spintronic applications \cite{Vobornik2011, Wang2016, Yabin2016, Yunkun2017, He2019, FERT2019}. For this purpose, it is interesting to open and manipulate a gap in the Dirac spectrum \cite{Dil2019ARPES}, either magnetically \cite{Ferreira2013TIQDots, Klinova2015TICmagneticfield, Calvo2017} or due to hybridization of edge/surface states in narrow systems \cite{Vayrynen2011, Dolcini2011EdgeInterferometry, Zhang2011, Ferreira2013TIQDots, Sternativo2014, Sternativo2014TICDisorder, Klinova2015TICmagneticfield}.

The use of different kinds of barriers and constrictions is the basis of the proposals of edge state Fabry-Pérot inteferometers \cite{Wu2010, Dolcini2011EdgeInterferometry, Romeo2012, Sternativo2014, Papaj2016TIC, Wu2017TICmemory, Nanclares2017, Karalic2020, osca2021fabryperot}, spin filters \cite{Krueckl2011TICSwitchspin, Klinova2015TICmagneticfield}, and Majorana fermion non-Abelian inteferometers \cite{Nilsson2010, Finck2014}.
Recent experimental progress on proximity coupling magnetism to topological insulator systems \cite{Bhattacharyya2020} supports the feasibility of engineering magnetic barriers.
It is also reassuring that Fabry-Pérot resonances have been experimentally observed in TI systems \cite{Finck2014,Finck2016,Calvo2017}.
Additionally, the TI spectrum can be simulated in photonic crystals to mimic these features \cite{Luo2011, Ma2015, Chang2016, Parappurath2020}.

\begin{figure}[t]
  \centering
  \includegraphics[width=1.0\columnwidth]{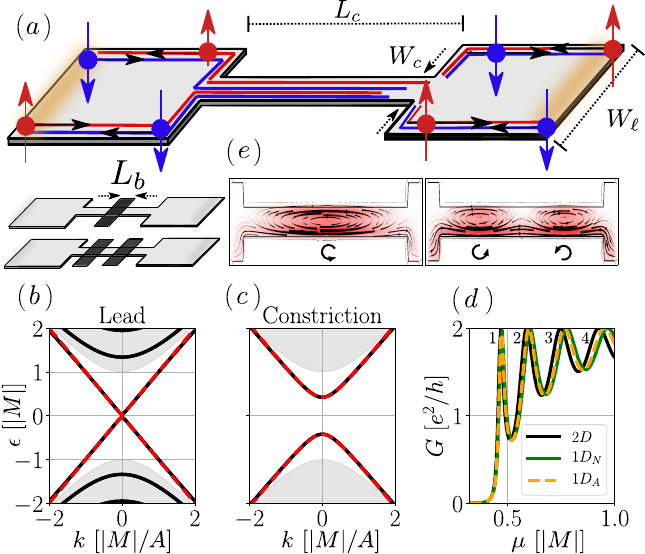}
  \caption{(a) Sketch of the TI constriction indicating its lengths and widths and the path of the helical edge states. The configuration with one or two barriers of length $L_b$ is also shown.
  (b) At the leads $W_\ell$ is large and the spectrum is massless, while (c) at the constriction a hybridization gap $\Delta$ opens if $W_c \sim \xi$ (penetration depth).
  In (b) and (c) the shaded regions mark the bulk bands, and the black (red) lines are the nanoribbon dispersion from the 2D BHZ (1D simplified) model.
  (d) The Fabry-Pérot resonant peaks in $G$ for the 2D and 1D models (analytical and numerical cases) match at low energies. (e) Spin-up component of the current densities within the constriction showing vortices at the first and second conduction peaks. The quantization index $n$ is indicated next to each peak.}
  \label{fig:model}
 \end{figure}

In this paper we investigate the formation of Fabry-Pérot resonances in the transport properties of narrow quasi-1D constrictions of TIs described by the BHZ model \cite{Bernevig2006BHZ}. We consider a geometry similar to that in the recent experiment in Ref.~\cite{Strunz2020Anomaly05}, but with a narrow enough channel to induce hybridization of the edge states within the constriction (see Fig. \ref{fig:model}). Consequently, reflections occur at both ends of the constriction, as it interfaces wider massless regions. The latter give rise to Fabry-Pérot resonances (FPRs) seen as conduction peaks [see Fig \ref{fig:model}(d)]. Interestingly, the reflection processes are associated with inter-edge scattering processes, yielding $n$ current vortices within the constriction for the $n$-th FPR.
We show how to manipulate these large FPR-driven conductance modulations by introducing one or two barriers defined by electrostatic (scalar) or magnetic contacts along the constriction, as shown in the insets in Fig. \ref{fig:model}(a). We find that a scalar contact simply shifts the peaks in energy, but its displacement depends on the intensity of the current vortex under the contact, which has $n-1$ destructive interference nodes within the constriction. Magnetic contacts split the FPR peaks as they hybridize into opposite spin polarizations. For the case of two barriers, the contacts can be set with parallel (P) or antiparallel (AP) magnetizations. For the P configuration the peaks split and polarize, while for the AP configuration it remains a single nonpolarized peak. As the peaks shift in energy when changed from the P to the AP configuration, the system shows large magnetoconductance variations.

The paper is organized as follows. In Sec.~\ref{sec:model} we present the model Hamiltonian and discuss the system geometry. In addition, we introduce a simple 1D model, whose analytical solution serves as a guide for our discussion. In Sec.~\ref{sec:results}, we study the main features of the Fabry-Pérot-induced current vortices and how to control the spin-resolved conductance by introducing magnetic barriers in the constriction.
All codes used in the numerical analysis of this work are available in the Supplemental Material \cite{SM}.
We present our conclusions in Sec.~\ref{sec:conclusions}.

\section{Model and parameters}
\label{sec:model}

We study the charge and spin transport properties of a two-probe 2D TI constriction of length $L_c$ and transverse width $W_c$ [see Fig. \ref{fig:model}(a)]. Semi-infinite leads of width $W_\ell$ are attached to the ends of the constriction. We consider that a small bias drives a net current flowing from the left $L$ (source) to the right $R$ (drain) lead.

For both the constriction and the leads, the TI is described by a $4\times4$ BHZ Hamiltonian \cite{Bernevig2006BHZ},
\begin{equation}
    H(k_x, k_y) = \begin{pmatrix}h(\bm{k}) & 0 \\ 0 & h^{*}(-\bm{k})\end{pmatrix},
    \label{eq:BHZ}
\end{equation}
where $h(\bm{k}) = \bm{k}\cdot\bm{\tau} + (m - \beta k^2)\tau_z$, $\bm{k} = (k_x, k_y)$, the Pauli matrices $\bm{\tau} = (\tau_x, \tau_y, \tau_z)$ act on the orbitals, and the $2\times2$ blocks of $H(k_x, k_y)$ refer to the spin-up and -down subspaces.
In order to emphasize the generality of our results and keep the notation compact, the block Hamiltonian $h(\bm{k})$ of the BHZ model \cite{Bernevig2006BHZ} is written in a dimensionless form. This is done by expressing the energy in units of the BHZ gap $M$, thus $m = M/|M| = \pm 1$, and using the penetration length $\xi = A/|M|$ as the length scale, yielding $\beta = |M|B/A^2$. For HgTe quantum wells the BHZ parameters (e.g., $A \approx 375$~meV~nm, $M \approx -10$~meV, and $B \approx -1120$~meV~nm$^2$) \cite{Konig2007QSHIHgTe, Bernevig2006BHZ} give $\beta \approx -0.1$. The usefulness of writing the BHZ model in a dimensionless form will become clear in Sec. \ref{sec:units}, where we address specific properties of different materials. The small value of $|\beta|$ guarantees that the bulk model is dominated by the gap and $k$-linear terms, while the $k^2$ term regularizes the lattice \cite{Messias2017FDP, Araujo2019BCDiracHam}. Time-reversal symmetry assures that pairs of helical states with opposite spin counter-propagate, that is, the spin-up (spin-down) current [red (blue) in Fig. \ref{fig:model}(a)] flows clockwise (anticlockwise) along the system edges.

For a semi-infinite TI layer, the helical states are exponentially localized at the system edge. The penetration depth \cite{Zhou2008finite-sizeQSH, Wada2011BiFilms} near the Dirac point ($\epsilon, k_x \sim 0$) is $\xi \approx 1$ (recall that $A/|M|$ is our length unit). We set $W_{\ell} \gg \xi$ to ensure that states localized at opposite lead edges do not hybridize, thus yielding a gapless band structure, as shown in Fig.~\ref{fig:model}(b). In distinction, we consider $W_c \gtrsim \xi$ leading to a gap opening at the constriction region [see Fig.~\ref{fig:model}(c)]. The hybridization gap at $k_{x}=0$ is $\Delta \propto e^{-W_{c}/\xi}$ \cite{shan2010TIsolutions}. A finite $\Delta$ is essential for the emergence of Fabry-Pérot resonances, since, for sufficiently low energies, $\Delta \neq 0$  implies  different carrier velocities ($v = \hbar^{-1}\partial \varepsilon/\partial k_x$) for the leads and the constriction.

To manipulate the FPR peaks, we also consider external top gates and/or magnetic contacts placed along the constriction, as shown in the insets in Fig. \ref{fig:model}(a). The corresponding Hamiltonian term acting on the contact region is
\begin{equation}
\label{eq:barriers}
    H_\nu = V_0 \sigma_\nu \otimes \tau_0,
\end{equation}
where $V_0$ defines the coupling intensity and $\nu = \{0, x, y, z\}$, with $\tau_0$ and $\sigma_0$ being the identities in orbital and spin spaces. Thus, a top gate (scalar) contact corresponds to $\nu = 0$ and magnetic contacts to $\nu = \{x, y, z\}$, depending on the magnetization direction. Here we discuss only the $\nu=\{0, x, z\}$ cases, since the $\nu=x$ and $\nu=y$ cases are qualitatively equivalent.

We calculate the Landauer conductance $G$ as a function of the chemical potential $\mu$ to characterize the Fabry-Pérot resonances. At zero temperature $G_{\sigma,\sigma'}(\mu) = (e^2/h)\mathcal{T}_{\sigma,\sigma'}(\mu)$, while for finite $T$,
\begin{align}
    G_{\sigma,\sigma'}(\mu) &= \dfrac{e^2}{h}\int  d\varepsilon
    \left(-\dfrac{\partial f}{\partial \varepsilon} \right)
    \mathcal{T}_{\sigma,\sigma'}(\varepsilon)
   ,
\end{align}
where $f(\varepsilon) = [1+e^{(\varepsilon-\mu)/k_BT}]^{-1}$ is the Fermi-Dirac distribution, and the indexes $(\sigma,\sigma')$ indicate the spin components injected at the source and collected at the drain, respectively.
By solving the system scattering matrix, one obtains
the transmission $\mathcal{T}_{\sigma,\sigma'}(\varepsilon)$ \cite{datta1997book}, as well as the spin-resolved local density of states and the electronic current density \cite{kwant, Santos2019}.

Next we present a simplified 1D model for the edge states, and later the full 2D model using the BHZ Hamiltonian, Eq.~\eqref{eq:BHZ}. By default, unless otherwise specified, we set our geometric parameters (in units of $A/|M|$) to $W_c = 1.5$, $W_\ell = 4$, $L_c = 15$, and $L_b = 1$ (see Fig.~\ref{fig:model}). These imply hybridization gaps of $\Delta_c \approx 0.4$ and $\Delta_\ell \approx 0.02 \ll \Delta_c$ (in units of $|M|$) within the constriction and leads, respectively.

\subsection{Simplified 1D model}
\label{sec:1dmodel}

To develop some insight into the main results, let us first analyze a simplified 1D model for the edge states given by the Hamiltonian
\begin{equation}
    H_{\rm 1D} =
    \delta (\sigma_0 \otimes \tau_z) +
    \alpha (\sigma_z \otimes \tau_x) k_x.
    \label{eq:1Dmodel}
\end{equation}
Here the basis is defined at $k_x=0$ as the symmetric $\ket{S,\sigma}$ and anti-symmetric $\ket{A,\sigma}$ combinations of edge states from opposite edges and with the same spin $\sigma$. These correspond to the hybridized edge states at $k_x=0$ with a hybridization gap $2|\delta|$. The Pauli matrices $\tau_\nu$ and $\sigma_\nu$ act on the orbital (A/S) and spin subspaces, respectively. The velocity term is $\alpha \approx A/|M| = 1$. In line with Figs. \ref{fig:model}(b) and \ref{fig:model}(c), we set $\delta = 0$ in the leads and $\delta = \Delta$ in the constriction. The finite $\Delta$ is extracted from the numerical solution of the 2D model described later. The dashed red lines in Figs. \ref{fig:model}(b) and \ref{fig:model}(c) show the energy dispersion of this 1D model in perfect agreement with the 2D model.

The simplicity of the model allows us to calculate the $S$ matrix analytically and write the transmission coefficient per spin channel [diagonal blocks of Eq.~\eqref{eq:1Dmodel}] as
\begin{align}
    \mathcal{T}_{\sigma,\sigma}(\varepsilon) &= \dfrac{2(\varepsilon^2 - \Delta^2)}{2\varepsilon^2-\Delta^2(1+\cos(\theta))},
    \label{eq:Trans}
\end{align}
with $\theta = (2L_c/\alpha)\sqrt{\varepsilon^2-\Delta^2}$. Solving for $\mathcal{T}_{\sigma,\sigma}(\varepsilon) = 1$, we find that the resonant peaks occur at the energies
\begin{equation}
    \varepsilon^\pm_n = \pm \sqrt{\left(\dfrac{\alpha}{L_c} n\pi\right)^2 + \Delta^2},
    \label{eq:Enpeak}
\end{equation}
which is simply the quantization of the energy dispersion of $H_{\rm 1D}$ with $k_x \rightarrow k_n = n\pi/L_c$, where $n=1,2,3,\cdots$. The expression above shows that the resonant peaks are restricted to energies above the hybridization gap, i.e., $|\varepsilon_n^\pm| > |\Delta|$. Assuming that isolated peaks have a Gaussian-like shape, a reasonable approximation except for the peak tails, their broadening $\gamma$ can be taken from a series expansion of $T(\varepsilon)$ for $\varepsilon \approx \varepsilon_n^{\pm}$, yielding for $L_c \gg 1$
\begin{align}
    \gamma_n \approx
    \dfrac{1}{\sqrt{2}}
    \Big(\dfrac{\alpha}{L_c}\Big)^3
    \Big( \dfrac{n \pi}{\Delta}\Big)^2.
\end{align}

\begin{figure}[t]
  \centering
  \includegraphics[width=1.0\columnwidth]{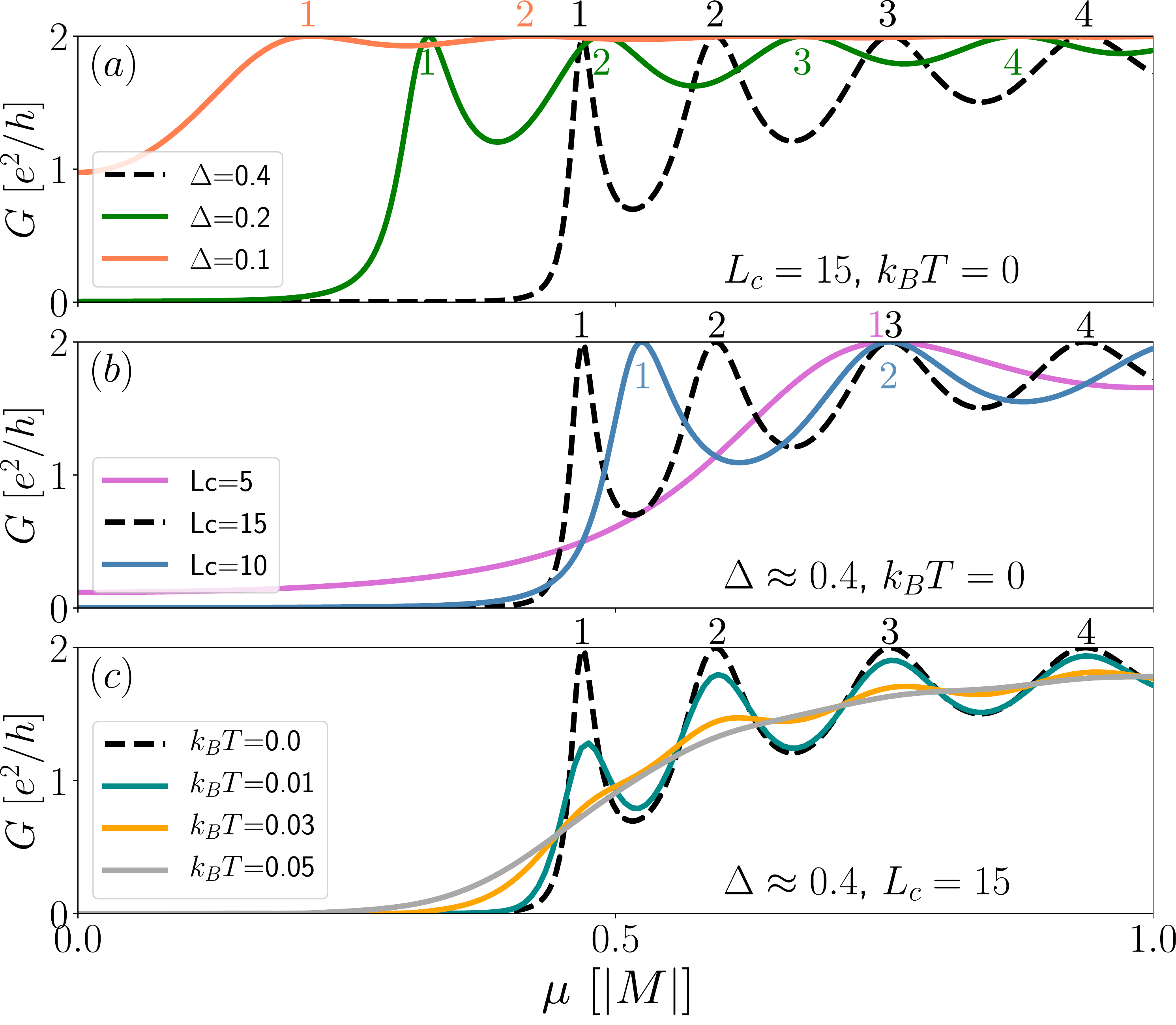}
  \caption{Total conductance $G$ at $T=0$ (a) for a fixed $L_{c}$ with a varying mass gap $\Delta$; and (b) for a fixed $\Delta$ with a varying $L_{c}$. (c) $G$ for finite $T$ showing the peaks smearing into a plateau as $k_BT$ increases. Dashed black lines represent our default parameters in the paper. The numbers next to a peak indicates its index $n$.}
  \label{fig:GvsParsTemp}
 \end{figure}

The simple expressions for $\varepsilon_n^\pm$ and $\gamma_n$ allow us to find model parameters to obtain isolated FPR peaks. For a given fixed $\Delta$, Eq.~\eqref{eq:Enpeak} shows that a large peak spacing requires a large $\alpha/L_c$, while a small broadening $\gamma_n$ is favored by a small  $\alpha/L_c$. The broadening $\gamma_n$ decreases also for increasing values of $\Delta$, but a large $\Delta$ shifts the FPR to larger energies $\varepsilon_n$, and the gap is $|\Delta| < 1$, since it is bounded by the bulk gap $|m| = 1$. Therefore, to get well-defined isolated FPR peaks one needs to balance these parameters [see Figs. \ref{fig:GvsParsTemp}(a) and \ref{fig:GvsParsTemp}(b)]. For large $L_c$ the peaks are pronounced and concentrated at low energies, while for small $L_c$ the peaks shift to larger energies and are suppressed. This occurs because at higher energies the effects of the hybridization gap vanish as the dispersion approaches $k$-linear behavior and the transmission becomes dominated by Klein tunneling \cite{klein1929reflexion, katsnelson2006chiral}. Similar effects occur for small $\Delta$. For our default set of parameters, presented in the previous section, we find $\gamma_1 = 0.01$ and $\varepsilon^\pm_2-\varepsilon^\pm_1 = 0.12$, which presents clear peaks in Fig.~\ref{fig:GvsParsTemp} (dashed black lines).

\subsection{Two-dimensional model}

Study of the transport properties of the BHZ model \cite{Bernevig2006BHZ} in the 2D constriction geometry shown in Fig. \ref{fig:model} requires numerical calculations. We compute the system conductance, local density of states, and current densities by implementing the BHZ Hamiltonian \cite{Bernevig2006BHZ}, Eq.~\eqref{eq:BHZ}, in Kwant \cite{kwant} using the finite differences approach on a square lattice, with lattice constant $a=0.2$ (for more details see, for instance, Ref.~\cite{Nanclares2017}). The codes developed for these calculations are available in the Supplemental Material \cite{SM}. The default set of parameters, presented above, is chosen to yield a large hybridization gap in the constriction region and well-defined peaks in $G$. Thus, since the penetration depth is $\xi = 1$, we choose $W_\ell = 4$ to guarantee a vanishing hybridization at the leads, and $W_c = 1.5$ to give an hybridization gap $\Delta = 0.4$ at the constriction. The constriction length is chosen as $L_c = 15$ to result in well-defined FPR peaks at low energies.

\section{Results}
\label{sec:results}

The results for the zero-temperature conductance obtained via Kwant \cite{kwant} simulations match well the 1D simplified model for energy ranges $|\varepsilon| \lesssim |m| = 1$, as shown in Fig. \ref{fig:model}(d).
For larger energies, extra channels are injected by the leads into the constriction and deviations from the 1D model are expected. Since our focus is on the low energy peaks, we can assume that the conductances obtained by both models always match.

At finite temperatures the conductance peaks are smeared, and $G$ tends towards a plateau shape [see Fig.~\ref{fig:GvsParsTemp}(c)]. Isolated resolved FPR peaks, which require  $\Delta\varepsilon_n \equiv |\varepsilon^\pm_{n+1}-\varepsilon^\pm_n| \gg k_BT$, demand an appropriate choice of $L_c$ and $\Delta$. For our default set of parameters $\Delta\varepsilon_1 \approx 0.12$ and $\gamma_1 \approx 0.01$. Indeed, Fig.~\ref{fig:GvsParsTemp}(c) shows that these FPR peaks would be clearly visible for $k_BT \lesssim 0.01$. 
Here, we neglect detrimental effects due to random spin-flip scattering processes and disorder \cite{Dolcini2011EdgeInterferometry, Sternativo2014, Sternativo2014TICDisorder}, which are likely to reduce the conductance peaks, but hardly qualitatively affect our results.

\subsection{Energy and length scales in real materials}
\label{sec:units}

To translate our dimensionless parameters into real materials let us first consider HgTe/CdTe quantum wells and a monolayer of GaBiCl$_2$ \cite{Li2015GaBiCl2}, which are quite extreme cases of the parameters. First, for HgTe/CdTe quantum wells the bulk gap is $M \approx -10$ meV \cite{Bernevig2006BHZ}, and the penetration depth is $\xi \approx 35$ nm. Therefore, for $W_c \approx 52.5$ nm and $L_c = 525$ nm, $\Delta \approx 4$ meV, $\Delta \varepsilon_1 \approx 1.2$ meV, and $\gamma_1 \approx 0.1$ meV, which tells us that the FPR peaks would be visible for $T \alt 1$ K. In contrast, GaBiCl$_2$ has the much larger gap of $M \approx -600$ meV, and, consequently, the much shorter penetration depth $\xi \approx 2$ nm. For these energy and length scales, the FPR peaks in GaBiCl$_2$ are clearly resolved for $T \alt 70$ K. In summary, the small bulk gap in HgTe/CdTe quantum wells leads to the requirement of low temperatures, while the short $\xi$ in GaBiCl$_2$ requires challenging narrow constrictions.

Therefore, the best platforms for experimental realization are those materials with an intermediate topological gap, which balances the parameters between the extreme cases discussed above. Fortunately, mid- and large-gap TIs have been intensively studied in 2D materials \cite{Wada2011BiFilms, Li2015GaBiCl2, Crisostomo2015IIIVHoneycombs, Wu2016NewFamilies, Wu2016EdgeStatesZrTe5, Zhang2017LargeGap, Levy2020EnhancedGapSL, ma2020spectroscopic} and there are several suitable candidates. To keep the discussion general, all results presented here are set in the general notation presented in Sec. \ref{sec:model} and the results are qualitatively valid for all 2D TI materials with the appropriate parameters as reported above.

\subsection{No barriers: Vortex formation}

\begin{figure}[t]
  \centering
  \includegraphics[width=1.0\columnwidth]{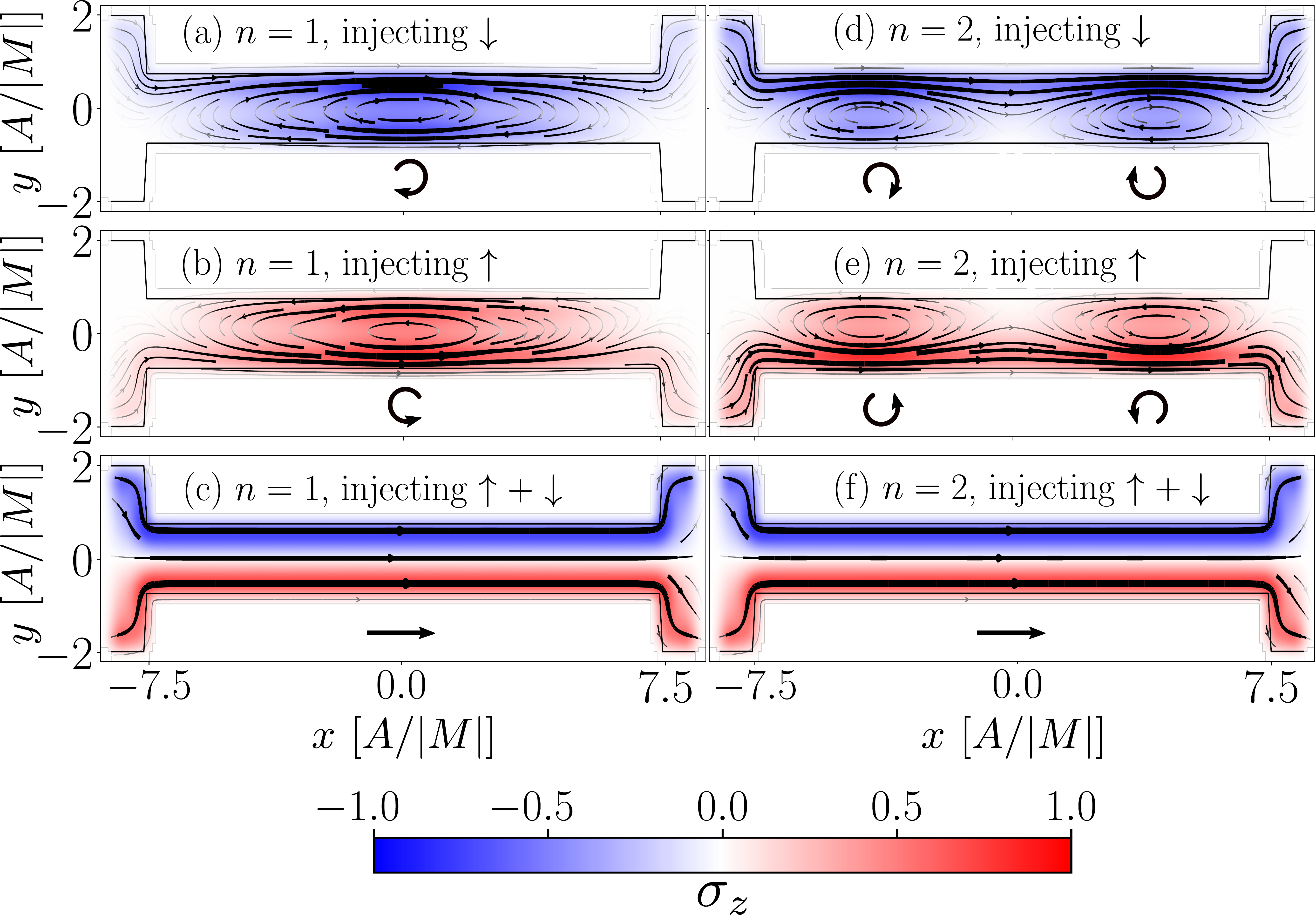}
  \caption{Current density (arrows) and spin $\sigma_z$ polarization [red (blue) for spin-up (spin-down)] across the constriction for the $n=1$ (left) and $n=2$ (right) FPR peaks in Fig. \ref{fig:GvsParsTemp} (dashed lines). The spin-up and spin-down channels show vortices rotating in opposite directions due to the helical nature of the TI. (c), (f) Adding both spin channels yields linear spin-polarized flow along the edges for both $n=1$ and $2$.}
  \label{fig:vortexformation}
\end{figure}

The FPR peaks in $G$ are associated with current vortices, as shown in Fig. \ref{fig:vortexformation}. The vortex formation is equivalent to that of stationary waves on a string, or the resonant modes on a double barrier resonant-tunneling diode \cite{TsuEsaki1973RTD}. The difference  is that here forward- and backward-moving electrons run along opposite edges of the constriction, leading to the formation of current vortices. The reflections occur at the ends of the constriction, since the hybridization gap $\Delta$ enforces that the velocity of the eigenstates at the constriction is smaller than at the leads, $\hbar |v|=|\partial\epsilon/\partial k| < \alpha$. The number $n$ of vortices for each spin channel is set by the nodes of destructive interference between the forward- and the backward-moving waves. The $n$-th peak in $G$ carries $n$ vortices, which also correspond to the quantization of $k_n = 2\pi/\lambda_n \rightarrow \lambda_n = 2L_{c}/n$.

Since, in a TI, electrons with opposite spins run in opposite directions at each edge, spin-up and -down modes form vortices circulating clockwise (spin-up) and counterclockwise (spin-down), respectively. If the system is time reversal symmetric, these vortices cancel each other in the total current density and electron flow becomes similar to that of a standard TI system, as shown in Figs. \ref{fig:vortexformation}(c) and \ref{fig:vortexformation}(f). To isolate a vortex, one needs to break the spin degeneracy. This can be achieved (i) by injecting a spin-polarized current, (ii) by applying an external magnetic field along $z$, or (iii) by applying a magnetic barrier that induces a field through a proximity effect, as shown below.

\subsection{Single barrier: peak splitting}

\begin{figure}[t]
  \centering
  \includegraphics[width=1.0\columnwidth]{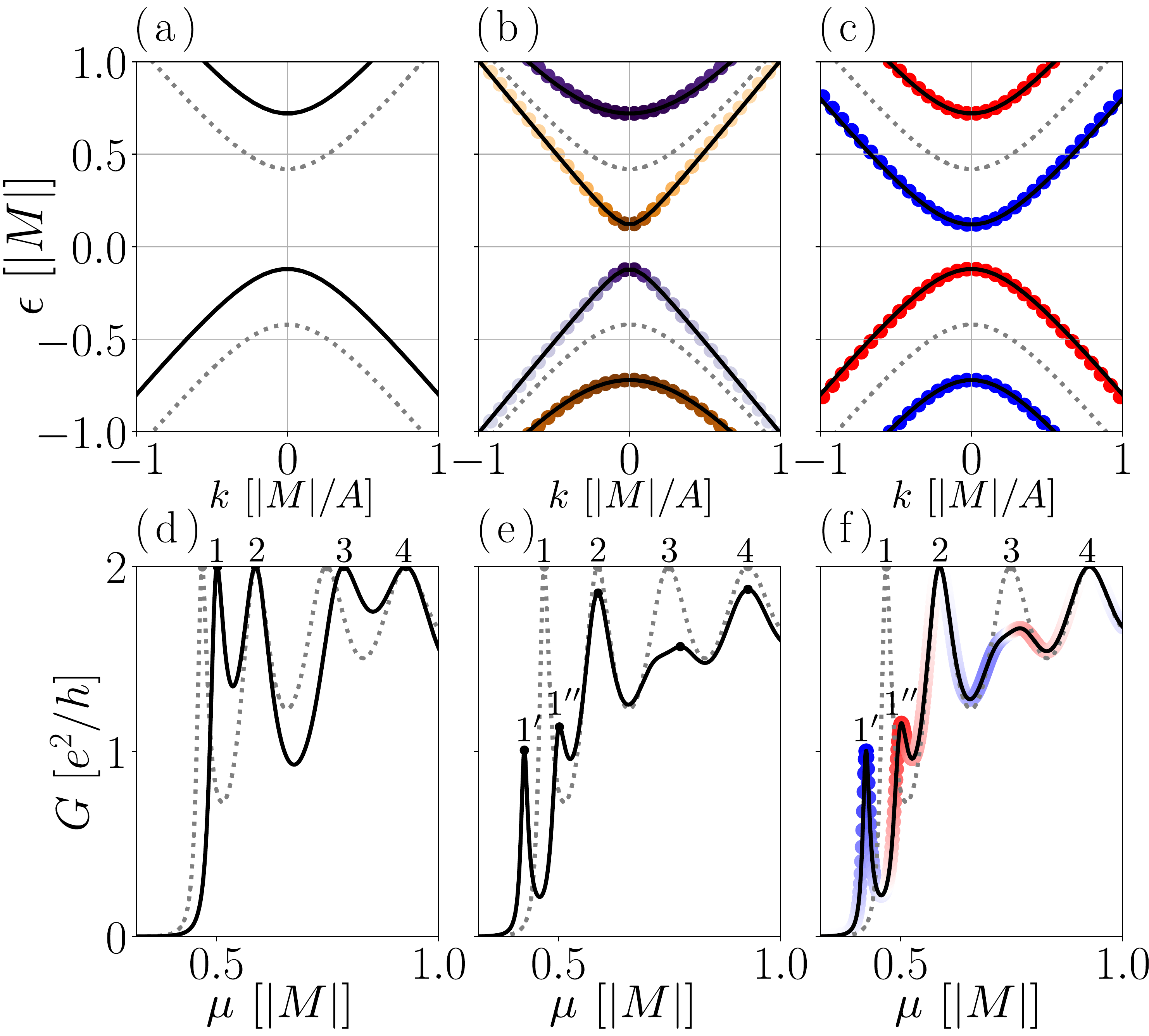}
  \caption{Energy spectrum of hybridized edge states within the barrier region with intensity $V_0 = 0.3$ for (a) a scalar ($\nu=0$) contact and for magnetic barriers polarized along (b) $\nu=x$ and (c) $\nu=z$. (d-f) Conductance $G(\mu)$ for a constriction with a single barrier corresponding to type $\nu$ in (a-c). The dashed gray line shows $G(\mu)$ for a constriction without barriers for reference. In (c) and (f) the color code indicates the spin $\sigma_z$ polarization [red (blue) for spin-up (spin-down)], while in (b) it indicates the $\sigma_x$ polarization [yellow for the expectation value $\mean{\sigma_x} \approx +1$ and blue for $\mean{\sigma_x} \approx -1$]. The $G$ peaks in (e) do not have a well defined polarization since $[H, \sigma_x] \neq 0$.}
  \label{fig:singlebarbands}
\end{figure}

The effects of a single barrier on the FPR peaks can be qualitatively understood as a perturbation. For instance, if a single and narrow barrier (width $L_b = 1$ and $V_0 = 0.3$) is placed at the center of the constriction, it matches a destructive interference node of the $n=2$ FPR peak [see Figs. \ref{fig:vortexformation}(d) and \ref{fig:vortexformation}(e)] and does not affect this peak (or other even-$n$ peaks), as shown in Figs. \ref{fig:singlebarbands}(d)-\ref{fig:singlebarbands}(f). For the other peaks, the effect depends on the type of barrier. Hereafter let us focus on the first $n=1$ peak, since it
shows the effects more clearly. As the broadening $\gamma_n$ increases with $n$ and the peak spacing is almost constant, FPR peaks with a larger $n$ tend to overlap.

An electrostatic barrier (scalar, $\nu=0$) modifies the system conductance in a trivial way: It rigid shifts the local band structure of the constriction [see Fig. \ref{fig:singlebarbands}(a)], shifting the FPR peak as well [Fig. \ref{fig:singlebarbands}(d)].
The resulting current density (not shown here) is the same as in the case of no barriers.
More interesting is the effect of a magnetic barrier polarized along $\nu=z$: It creates a local Zeeman field that splits the bands [see Fig.~\ref{fig:singlebarbands}(c)]. Hence,
the $n=1$ FP conductance peak splits into spin-up and -down resonances, indicated by $n=1^\prime$ and $1^{\prime\prime}$ in Fig. \ref{fig:singlebarbands}(f). In this case, due to the breaking of spin degeneracy, the current vortices do appear in the total current density.

\begin{figure}[t]
  \centering
  \includegraphics[width=\columnwidth]{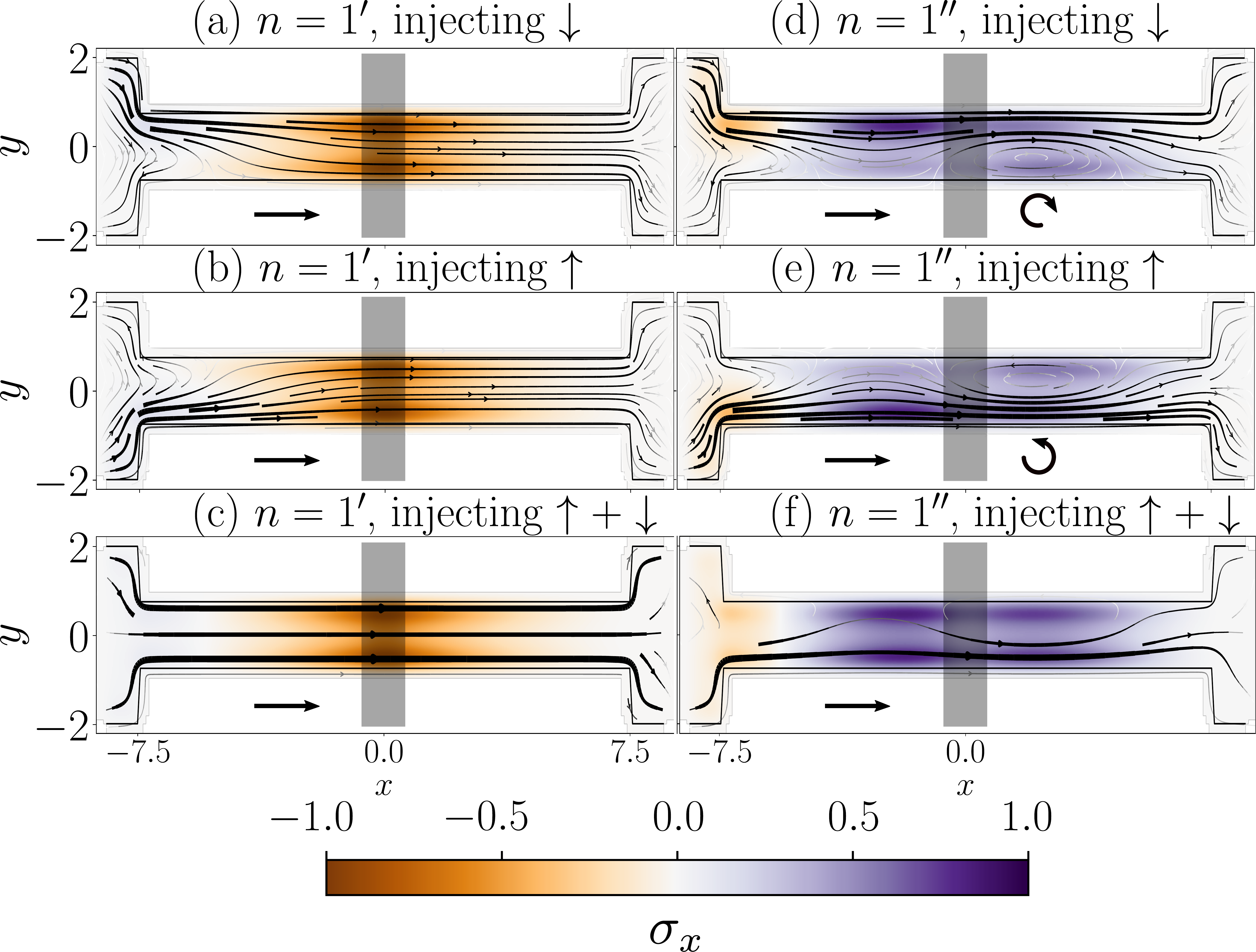}
  \caption{Current density (arrows) and spin $\sigma_x$ polarization across a constriction with a single $\nu=x$ barrier at the center (shaded region). The left (right) panels correspond to the first (second) $G$ peak in Fig. \ref{fig:singlebarbands}(e), with the source injecting only spin-up and -down channels on top, and their sum at the bottom. Near the barrier the current is approximately $\sigma_x$ polarized, but it loses polarization as it approaches the leads.}
  \label{fig:1barCurDens}
\end{figure}

For a magnetic barrier polarized along $\nu=x$, the local bands hybridize with spin polarization $\mean{\sigma_x} \approx \pm 1$, as shown in Fig. \ref{fig:singlebarbands}(b).
The conductance peaks in Fig. \ref{fig:singlebarbands}(e) do not have well-defined polarization, since the current flows through regions with and without barriers along the constriction and $[H, \sigma_x] \neq 0$. Consequently, the relative phase between the spin-up and the spin-down components on electrons arriving at the drain strongly depends on the current path and oscillates along the $(\sigma_x,\sigma_y)$ directions. In this case the vortex is destroyed by the hybridization and reflections at the barrier (see Fig. \ref{fig:1barCurDens}). Interestingly, around the barriers the current becomes spin $\sigma_x$ polarized with opposite polarizations for the first and second $G$ peaks, which is a consequence of the hybridization of the bands imposed by the barrier [Figs. \ref{fig:singlebarbands}(b) and \ref{fig:singlebarbands}(e)]. However, the current density loses this polarization as it approaches the leads, where the eigenstates are $\sigma_z$ quantized.

\subsection{Two barriers: Magnetoconductance}

We consider now a set up with two barriers placed within the constriction. The corresponding Hamiltonian consists of two terms like $H_\nu$ in Eq.~\eqref{eq:barriers} with $V_0\rightarrow V_1$ and $V_0\rightarrow V_2$ set in ``parallel''~(P; $V_2=V_1$) or ``antiparallel''~(AP; $V_2=-V_1$) configurations. For this setup we set $V_0=0.2$. The barriers are placed at 1/3 and 2/3 of the constriction length, as shown in the inset in Fig. \ref{fig:model}. These positions  match with the nodes of the $n=3$ FPR, thus barely affecting the third $G$ peak, $n=3$.
As above, we focus the discussion on the first peak.

\begin{figure}[t]
  \centering
  \includegraphics[width=\columnwidth]{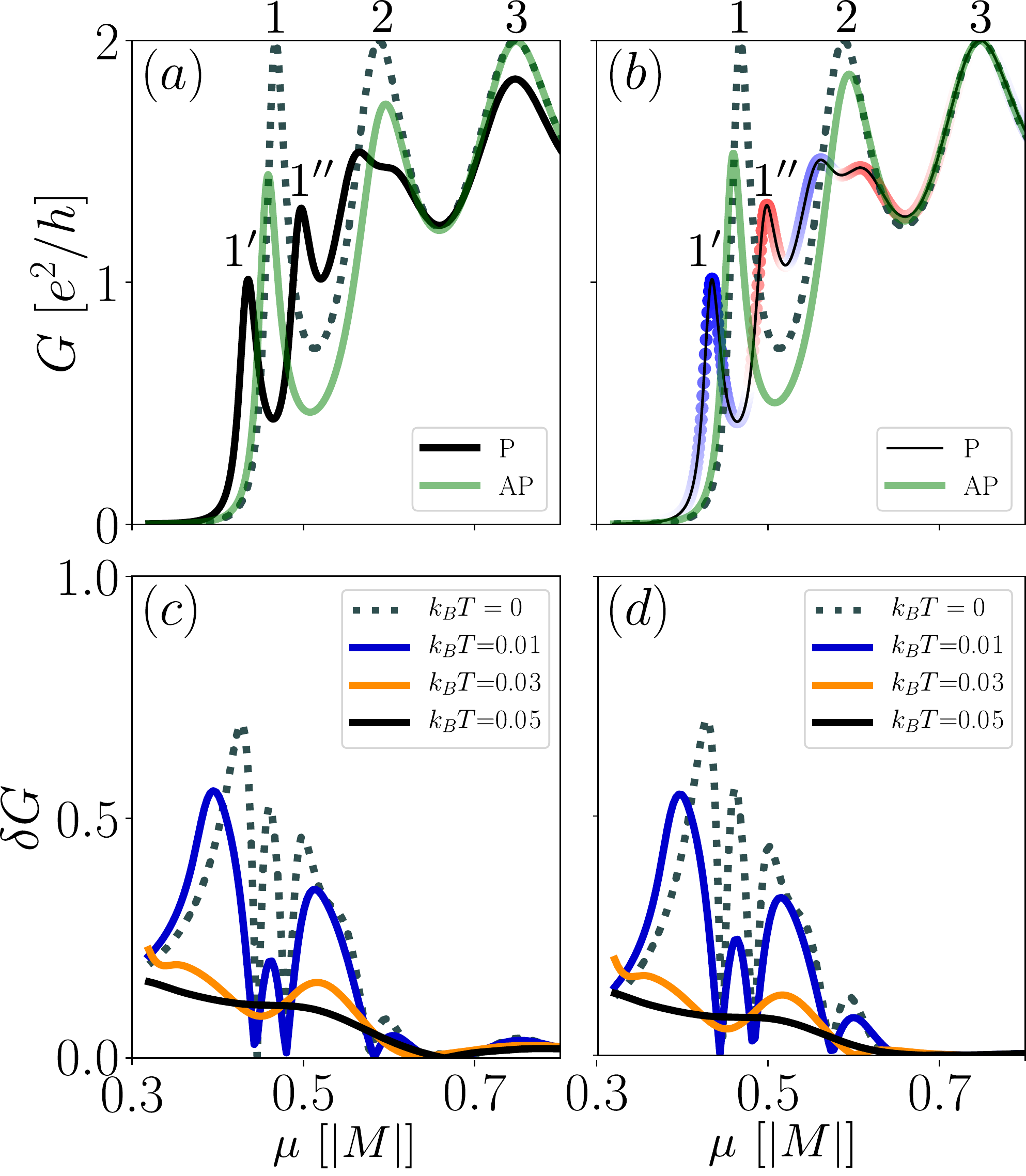}
  \caption{Conductance $G$ for the case of two magnetic barriers in the P and AP configurations polarized along (a) $\nu=x$ and (b) $\nu=z$. Dashed gray lines show $G$ without barriers as a reference. The color code in (b) indicates the spin $\sigma_z$ polarization. (c, d) The lower panels show the magnetoconductance $\delta G$ [see Eq. \eqref{eq:magnG}] for different temperatures for $\nu$ matching (a) and (b).}
  \label{fig:2barsGMR}
\end{figure}

In all cases the P configuration only enhances either the peak shifts ($\nu=\{0,z\})$ or the hybridization ($\nu=x$) already seen in the single-barrier case (cf. Fig. \ref{fig:singlebarbands} and Fig. \ref{fig:2barsGMR}). In the AP configuration the peaks do not split, but only slightly shift and weaken due to reflections induced by the magnetic barriers, as shown in Fig. \ref{fig:2barsGMR}.

The large conductance difference between the P and the AP magnetic barrier configurations leads to a non-trivial magnetoconductance, which we define as
\begin{equation}
    \delta G_\nu (\mu) = \left| \dfrac{G_P^\nu (\mu)- G_{AP}^\nu (\mu)}{G_P^\nu (\mu) + G_{AP}^\nu (\mu)} \right|,
    \label{eq:magnG}
\end{equation}
where $G_P^\nu$ and $G_{AP}^\nu$ are the total conductances for the P and AP configurations of the double-barrier of type $\nu=\{x,z\}$. As shown in Fig. \ref{fig:2barsGMR}, at low temperatures $\delta G$ reaches large values over broad ranges of $\mu$. In Figs. \ref{fig:2barsGMR}(a) and \ref{fig:2barsGMR}(b) the FPR peak of the AP configurations lies within the hybridized peaks of the P configuration, yielding a strong magnetoconductance, $\delta G \sim 0.6$ [Figs. \ref{fig:2barsGMR}(c) and \ref{fig:2barsGMR}(d)], with a large peak-to-valley ratio of $\sim 1.5:0.5$ at zero temperature. At higher temperatures $\delta G$ is reduced but remains $\sim 0.25$ at $k_BT = 0.03$. Interestingly, while an increase in $k_BT$ lowers $\delta G$, it also broadens the peaks, making the overall effect more robust against uncertainties in $\mu$.

\begin{figure*}[t]
  \centering
  \includegraphics[width=\textwidth]{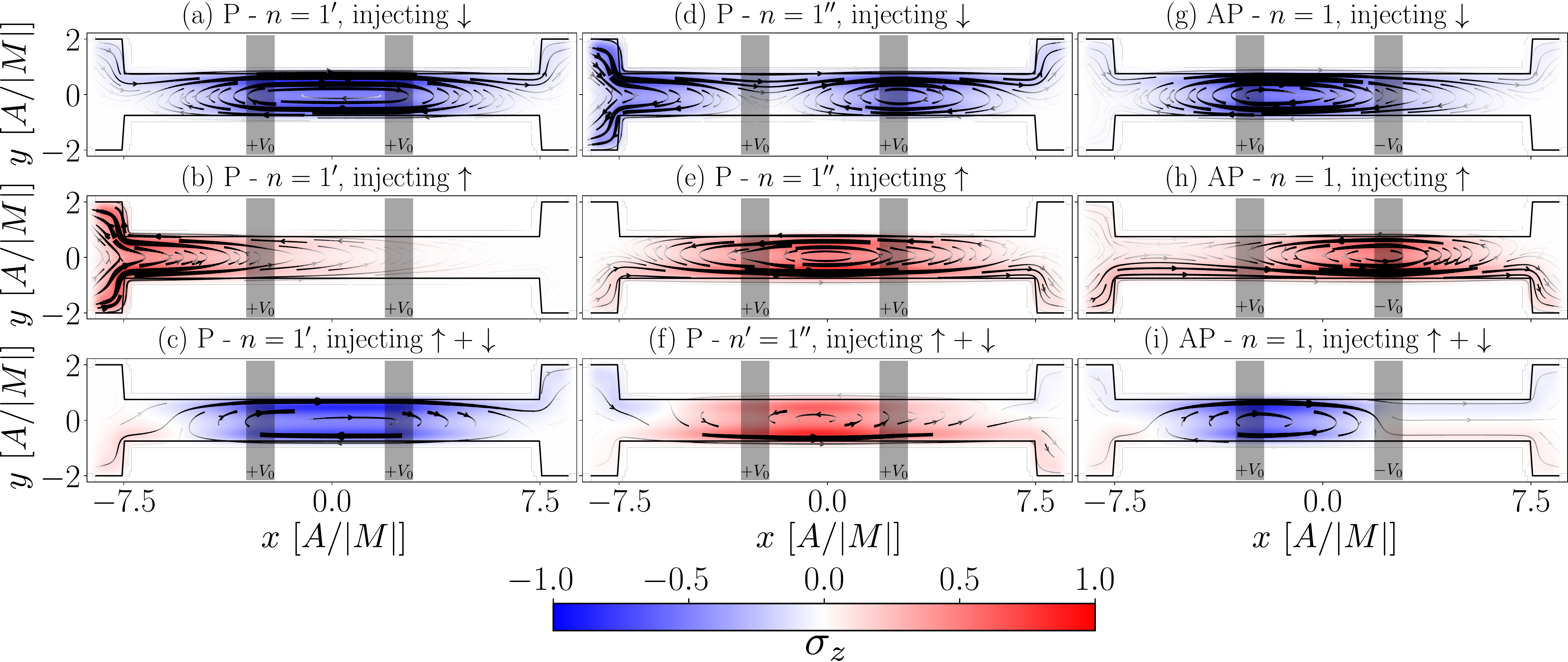}
  \caption{Current density (arrows) and spin $\sigma_z$ polarization across two $\nu=z$ barriers (shaded regions centered at $x=\pm 2.5$). (a-c) In the first $G$ peak in Fig. \ref{fig:2barsGMR}(b) for the P configuration, the spin-down (blue) channel is resonant, while the spin-up (red) channel reflects at the barrier, leading to a total current with a net spin-down polarization. (d-f) The second peak in the P configuration has the opposite spin polarization. (g-i) In the AP configuration, the spin-down channel reflects on the second barrier and the resonant vortex forms in the first section of the constriction, while the spin-up resonance occurs in the second half.}
  \label{fig:2ZbarsCurrDens}
\end{figure*}

The current densities for the double-barrier case polarized along $\nu=z$ are shown in Fig. \ref{fig:2ZbarsCurrDens}. In the P configuration the current densities correspond to the peaks $n=1'$ and $1''$, while those for AP barriers represent the $n=1$
single peak [see Fig. \ref{fig:2barsGMR}(b)]. For the $n=1'$ P peak [Figs. \ref{fig:2ZbarsCurrDens}(a)-\ref{fig:2ZbarsCurrDens}(c)] the spin-down component forms a resonant vortex, while the spin-up component is reflected at the barriers, leading to a total current with a spin-down polarized vortex. For the $n=1''$ P peak [Figs. \ref{fig:2ZbarsCurrDens}(d)-\ref{fig:2ZbarsCurrDens}(f)] the opposite spin component dominates since the peaks correspond to a local Zeeman splitting within the barriers. However, due to its higher energy, the reflection of the spin-down  channel in Fig. \ref{fig:2ZbarsCurrDens}(d) is not as intense as that of the spin up channel, shown in Fig \ref{fig:2ZbarsCurrDens}(b). For the AP configuration, depicted in Figs. \ref{fig:2ZbarsCurrDens}(g)-\ref{fig:2ZbarsCurrDens}(i), the local Zeeman splitting of each barrier is reversed, with the first (second) barrier concentrating a spin-down (spin-up) resonant vortex. The total current density shown in Fig. \ref{fig:2ZbarsCurrDens}(i) favors the spin-down polarization of the first barrier, since the spin-up channel partially reflects on the first barrier before forming the resonant vortex near the second barrier.

\subsection{Vortex detection}

Here, we propose an indirect strategy to assess the local current densities analyzed in this paper by invoking a 2D counterpart of the topological magnetoelectric effect \cite{Qi2008, Dziom2017, Liu2020}, that is, the current vortices induce a magnetic field perpendicular to the constriction plane that can be experimentally measured.

Let us first discuss the case where the currents are spin polarized, due, for instance, to the injection by ferromagnetic terminals. In this case, for $\mu$ values corresponding to the $n$-th conductance peak one expects $n$ vortices within the constriction, as shown in Fig.~\ref{fig:vortexformation}. At the center of each vortex, we can estimate the magnetic field using the Biot-Savart law for an elliptical current loop, which gives \cite{Romero2021}
\begin{align}
    B_z &= \dfrac{\mu I}{\pi b} E\big(1-(b/a)^2\big),
\end{align}
where $a$ and $b$ are the lengths of the ellipse semi-major and semi-minor axes, $I$ is the current, $\mu$ is the magnetic permeability, and $E(\cdots)$ is the complete elliptic integral of the second kind. For the $n=1$ vortex [\textit{e.g.}, in Fig.~\ref{fig:vortexformation}(a)] we have $a=L_c/2$ and $b=W_c/2$. Assuming a typical $I=1$~$\mu$A, $\mu=\mu_0$ (vacuum permeability), and for the parameters stated in Sec. \ref{sec:units}, we estimate $B_z \approx 0.27$~mT for GaBiCl$_2$ and $B_z = 0.015$~mT for HgTe. For other $G$ peaks with $n>1$, each vortex now has a major semiaxis $a=L_c/2n$ and the estimates for $B_z$ slightly increase. More importantly, $B_z$ will be modulated across the constriction, being maximum at the center of each vortex. Its position dependence allows one, in principle, to reconstruct the  local current profile. In contrast, for the non-spin-polarized injection, as in Figs.~\ref{fig:vortexformation}(c) and \ref{fig:vortexformation}(f), the magnetic field at the center of the constriction is expected to be $B_z = 0$. Therefore, both the intensity and the modulation of the field along the constriction are signatures of the successful injection of spin-polarized currents.

For nonpolarized spin injection, one can introduce magnetic barriers in the constriction to generate spin-polarized vortex currents. See, for instance, Fig.~\ref{fig:2ZbarsCurrDens}. Using the ideas described above, the current profiles can be obtained from $B(x,y)$.

\section{Conclusions}
\label{sec:conclusions}

We have investigated the characteristics of the conductance across TI constrictions, which show Fabry-Pérot resonances due to the hybridization of edge states and reflections at the ends of the constriction. The dynamics of a packet moving forward along one edge and reflecting through the opposite edge leads to vortices in the current density. These current vortices induce out-of-plane magnetic fields ${\bf B} = B_z \hat{{\bf e}}_z$, which we identify as a 2D counterpart of the magnetoelectric effect \cite{Qi2008, Dziom2017, Liu2020}. The modulation of $B_z$ along the constriction, being maximum at the center of each of the $n$ vortices, is a signature of successful injection of a spin-polarized current, and it might be useful in the design of magnetic memories \cite{Manipatruni2019}. We show that to get well-defined FPR peaks it is desirable to have a topological gap larger than the temperature and an edge state penetration depth within the range of lithographically producing the sample constriction. Since the gap and $\xi$ are inversely proportional to each other, it is best to work with intermediate-gap TI materials. The proposal might also serve as an interesting magnetoconducance device if two magnetic barriers (produced, for instance,  via proximity effect) are able to switch between parallel and antiparallel configurations.

The model and results discussed here were obtained using the single-particle picture, as is standard for most 2D TI materials. Interestingly, recent experimental data \cite{Strunz2020Anomaly05} on a quantum point contact made by lateral constrictions in HgTe/CdTe quantum wells suggest that electronic interactions can lead to a 0.5 anomaly. This experimental finding, which resembles the 0.7 anomaly observed in ordinary semiconductor quantum point contacts, is not well understood yet. In distinction, the constrictions considered in this paper are much narrower and much longer than the quantum point contact in Ref.~\cite{Strunz2020Anomaly05} and are more similar to mesoscopic semiconductor quantum
wires, which are nicely understood in terms of the single-particle picture.

\section{Acknowledgments}

This work was financially supported by the the Brazilian funding agencies CNPq, CAPES, FAPERJ, and FAPEMIG.

\bibliography{constriction}
\end{document}